\documentclass[epj]{svjour}
\usepackage[dvips]{graphicx}
\usepackage{array}
\usepackage{amssymb}
\usepackage{amsmath}
\usepackage{hhline}
\usepackage{longtable}
\begin{document}
\title{Evolving pseudofractal networks}
\author{Zhongzhi Zhang\inst{1,2} \and Shuigeng Zhou\inst{1,2} \and Lichao Chen\inst{1,2}}                     % Do not remove
\offprints{zhangzz@fudan.edu.cn (Zhongzhi Zhang), sgzhou@fudan.edu.cn (Shuigeng Zhou)}          % Insert a name or remove this line

\institute{Department of Computer Science and Engineering, Fudan
University, Shanghai 200433, China \and Shanghai Key Lab of
Intelligent Information Processing, Fudan University, Shanghai
200433, China}
\date{Received: date / Revised version: date}

\abstract{We present a family of scale-free network model consisting
of cliques, which is established by a simple recursive algorithm. We
investigate the networks both analytically and numerically. The
obtained analytical solutions show that the networks follow a
power-law degree distribution, with degree exponent continuously
tuned between 2 and 3. The exact expression of clustering
coefficient is also provided for the networks. Furthermore, the
investigation of the average path length reveals that the networks
possess small-world feature. Interestingly, we find that a special
case of our model can be mapped into the Yule process.
 % end of PACS codes
} %end of abstract

\maketitle

%%%%%%%%%%%%%%%%%%%%%%%%%%%%%%%%%%%%%%%%%%%%%%%%%%%%%%%%%%%%%%%%%%%%
\section{Introduction}
Over the last few years, it has been suggested that a lot of social,
technological, biological, and information networks share the
following three striking statistical characteristics
\cite{AlBa02,DoMe02,Ne03,BoLaMoChHw06}: power-law degree
distribution \cite{BaAl99}, high clustering coefficient
\cite{WaSt98}, and small average path length (APL). Power-law degree
distribution indicates that the majority of nodes (vertices) in such
networks have only a few connections to other nodes, whereas some
nodes are connected to many other nodes in the network. Large
clustering coefficient implies that nodes having a common neighbor
are far more likely to be linked to each other than are two nodes
selected randomly. Short APL shows that the expected number of links
(edges) needed to pass from one arbitrarily selected node to another
one is low, that is, APL grows logarithmically with the number of
nodes or slower.

In order to mimic such complex real-life systems, a wide variety of
models have been proposed~\cite{AlBa02,DoMe02,Ne03,BoLaMoChHw06},
among which the most well-known successful attempts are the Watts
and Strogatz's (WS) small-world network model \cite{WaSt98} and
Barab\'asi and Albert's (BA) scale-free network model \cite{BaAl99},
which have attracted an exceptional amount of attention from a wide
circle of researchers and started an avalanche of research on the
models of systems within the physics community. After that, a
considerable number of other models and mechanisms, which may
represent processes more realistically taking place in real-life
systems, have been developed. These mainly include nonlinear
preferential attachment~\cite{KaReLe00}, initial
attractiveness~\cite{DoMeSa00}, edge rewiring~\cite{AlBa00} and
removal~\cite{DoMe00b}, aging and cost~\cite{AmScBaSt00},
competitive dynamics~\cite{BiBa01}, duplication~\cite{ChLuDeGa03},
weight~\cite{WWX1,ZhZhFaGuZh07}, geographical
constraint~\cite{RoCoAvHa02,ZhRoCo05a}, Apollonian
packing~\cite{AnHeAnSi05,DoMa05,ZhCoFeRo05,ZhYaWa05,ZhRoCo05,ZhRoZh06}
and so forth. Today, modeling complex systems with small-world and
scale-free characteristics is still an important issue.

Recently, Dorogovtsev, Goltsev, and Mendes have demonstrated that
scale-free behavior and small-world effect can be excellently
modeled by using pure mathematical objects and methods to construct
a deterministic graph~\cite{DoGoMe02}, called pseudofractal
scale-free network (PSFN) which was extended by Comellas \emph{et.
al.} in \cite{CoFeRa04}. PSFN has drawn much attention from the
scientific community, many dynamical processes taking place in PSFN
have been intensively studied, including synchronization
\cite{LiGaHe04}, diffusion \cite{Bobe05}, and opinion formation
\cite{GoSoHe06}. The PSFN is a regular deterministic network in a
certain sense without statistical mechanics for consideration. In
\cite {DoMeSa01}, Dorogovtsev, Mendes and Samukhin proposed a random
growing network, which we call random pseudofractal scale-free
network (RPSFN). The PSFN and RPSFN may provide valuable insight
into some particular real-life networks.

In this paper, we propose a general scenario for constructing
evolving pseudofractal networks (EPNs) governed by three parameters
$m$, $p$, and $q$, which control the relevant network
characteristics. The EPN unifies the PSFN and RPSFN to the same
framework, i.e. the PSFN and RPSFN are special cases of EPN. In
addition to PSFN and RPSFN, the EPN also includes many other models
as its particular cases. More interestingly, one particular case of
EPNs can be mapped into the Yule process. The growing EPNs are
composed of cliques, and result in a power-law degree distribution
with degree exponent changeable between 2 and 3, a very large
clustering coefficient, and a small-world feature.

%%%%%%%%%%%%%%%%%%%%%%%%%%%%%%%%%%%%%%%%%%%%%%%%%%%%%%%%%%%%%%%%%%%
%%  Network construction       %%%%%%%%%%%%%%%%%
%%%%%%%%%%%%%%%%%%%%%%%%%%%%%%%%%%%%%%%%%%%%%%%%%%%%%%%%%%%%%%%%%%%
\section{Network construction}
\begin{figure}
\begin{center}
\includegraphics[width=8cm]{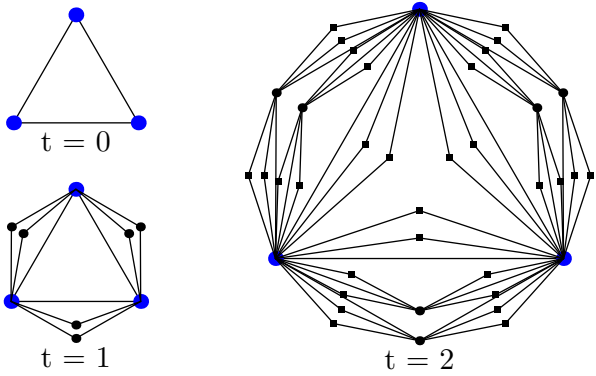}
\caption{Illustration of a deterministically growing network in the
case of $m=2$, $p=1$, and  $q=2$, showing the first three steps of
growing process. } \label{network}
\end{center}
\end{figure}

We construct the networks in a recursive manner and denote the
networks after $t$ generations by $Q(q,t)$, $q\geq 2, t\geq 0$. Then
the network construction process is as follows: For $t=0$, $Q(q,0)$
is a complete graph $K_{q+1}$ (or $(q+1)$-clique). For $t\geq 1$,
$Q(q,t)$ is obtained from $Q(q,t-1)$. For each of the existing
subgraphs of $Q(q,t-1)$ that is isomorphic to a $q-$clique, with
probability $p$ $(0<p\leq 1)$, $m$ ($m$ is a positive integer) new
vertices are created, and each is connected to all the vertices of
this subgraph. The growing process is repeated until the network
reaches a desired size. Figure~\ref{network} shows the network
growing process for a particular case of $m=2$, $p=1$, and $q=2$.
%%%%%%%%%%%%%%%%%%%%%%%%%%%%%%%%%%%%%%%%%%%%%%%%%%%%%%%%%%
% Figure  1
%%%%%%%%%%%%%%%%%%%%%%%%%%%%%%%%%%%%%%%%%%%%%%%%%%%%%%%%%%
%%%%%%%%%%%%%%%%%%%%%%%%%%%%%%%%%%%%%%%%%%%%%%%%%%%%%%%%%%

There are at least five limiting cases of our model listed below.
(i) When $m=1$, $p=1$, and $q=2$, the networks are exactly the same
as the pseudofractal scale-free network (PSFN)~\cite{DoGoMe02}. (ii)
When $m=1$, $p\rightarrow0$ (but $p \neq0$), and $q=2$, our model is
reduced to the random pseudofractal scale-free network
(RPSFN)~\cite{DoMeSa01}. (iii) When
 $m=1$, $0<p\leq 1$, and $q=2$, our networks coincide with the
stochastically growing scale-free network described in~\cite{Do03}.
(iv) When $m=1$, $p=1$, and $q\geq 2$, our networks reduce to the
recursive graphs discussed in~\cite{CoFeRa04}. (v) When  $p=1$ and
$q\geq 2$, our networks turn out to be the graphs introduced
in~\cite{ZhRoZh07}. Thus, varying parameters $m$, $p$, and $q$, we
can study many crossovers between these limiting cases.

Next we compute the numbers of nodes and links in $Q(q,t)$. Let
$L_v(t)$, $L_e(t)$ and $K_{q,t}$ be the numbers of vertices, edges
and $q$-cliques created at step $t$, respectively. Note that the
addition of each new node leads to $q$ new $q$-cliques and $q$ new
edges. So, we have $L_e(t)=K_{q,t}=qL_v(t)$. Then, at step 1, we add
expected $L_v(1)=mp(q+1)$ new nodes and $L_e(1)=mpq(q+1)$ new edges
to $Q(q,0)$. After simple calculations, one can obtain that at
$t_i$($t_i>1$) the numbers of newly born nodes and edges are
$L_v(t_i)=mp(q+1)(1+mpq)^{t_i-1}$ and
$L_e(t_i)=mpq(q+1)(1+mpq)^{t_i-1}$, respectively. Thus the average
number of total nodes $N_t$ and edges $E_t$ present at step $t$ is
\begin{eqnarray}\label{Nt1}
N_t=\sum_{t_i=0}^{t}L_v(t_i)=\frac{(q+1)[(mpq+1)^{t}+q-1]}{q}
\end{eqnarray}
and
\begin{eqnarray}\label{Et1}
E_t=\sum_{t_i=0}^{t}L_e(t_i)=\frac{(q+1)[2(mpq+1)^{t}+(q-2)]}{2},
\end{eqnarray}
respectively. So for large $t$, The average degree $\overline{k}_t=
\frac{2E_t}{N_t}$ is approximately $2q$.

\section{Topological properties}
Topology properties are of fundamental significance to understand
the complex dynamics of real-life systems. Here we focus on three
important characteristics: degree distribution, clustering
coefficient, and average path length, which are determined by the
tunable model parameters $m$, $p$, and $q$.

\subsection{Degree distribution}
Degree distribution is one of the most important statistical
characteristics of a network. Firstly, we follow the method that was
introduced in \cite{ZhCoFeRo05,ZhRoZh06} for the calculation of
degree distribution for the general case; then, we use the
rate-equation approach~\cite{KaRe01} to get the degree distribution
for some limiting cases.

 \subsubsection{General case}
\begin{figure}
\begin{center}
\includegraphics[width=9cm]{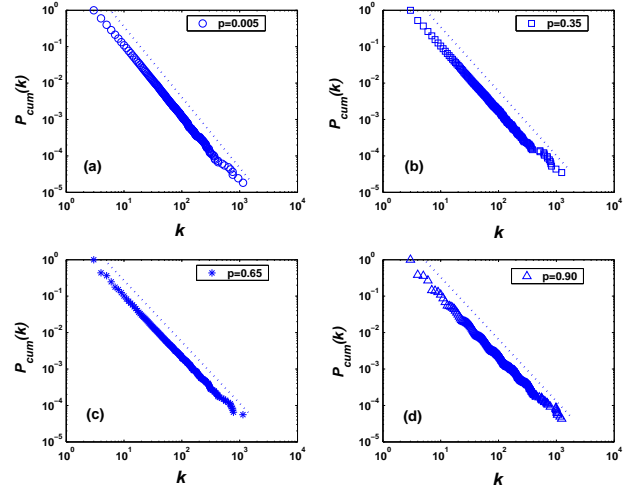}
\caption{The cumulative degree distribution $P_{cum}(k)$ at various
$p$ values for the case of $m=1$ and $q=2$. The circles (a), squares
(b), stars (c), and triangles (d) denote the simulation results for
networks with different evolutionary steps $t=1350$, $t=25$, $t=16$,
and $t=13$, respectively. The four straight lines are the
theoretical results of $\gamma(m,p,q)$ as provided by equation
(\ref{gamma}). All data are from the average of 50 independent
runs.} \label{Fig3}
\end{center}
\end{figure}

 When a new node $i$ is added to the
networks at step $t_i$, it has degree $q$ and forms $q$ $q$-cliques.
Let $L_q(i,t)$ be the number of $q$-cliques at step $t$ that will
possibly create new nodes connected to the node $i$ at step $t+1$.
At step $t_i$, $L_q(i, t_i)=q$. By construction, we can see that in
the subsequent steps each new neighbor of $i$ generates $q-1$ new
$q$-cliques with $i$ as one node of them.  Then at step $t_i+1$,
there are $mpq$ new nodes which forms $mpq(q-1)$ new $q$-cliques
containing $i$.  Let $k_i(t)$ be the degree of $i$ at step $t$. We
can easily find following relations for $t>t_i+1$:
\begin{equation}
\Delta k_i(t)=k_i(t)-k_i(t-1)=mpL_q(i,t-1)
\end{equation}
and
\begin{equation}
L_q(i,t)=L_q(i,t-1)+(q-1)\Delta k_i(t).
\end{equation}
From the above two equations, we can derive: $L_q(i, t+1)= L_q(i,
t)[1+mp(q-1)]$. Since $L_q(i, t_i)=q$, we have $L_q(i,
t)=q[1+mp(q-1)]^{t-t_i}$ and $\Delta
k_i(t)=mpq[1+mp(q-1)]^{t-t_i-1}$. Then the degree $ k_i(t)$ of node
$i$ at time $t$ is
\begin{eqnarray} \label{Ki1}
k_i(t)&=k_i(t_i)+\sum_{t_h=t_i+1}^{t}{\Delta k_i(t_h)}\nonumber\\
&=q\left(\frac{[1+mp(q-1)]^{t-t_i}+q-2}{q-1}\right).
\end{eqnarray}
Since the degree of each node has been obtained explicitly as in
Eq.~(\ref{Ki1}), we can get the degree distribution via its
cumulative distribution \cite{Ne03}, i.e., $P_{cum}(k) \equiv
\frac{1}{N_t}\sum_{k^\prime \geq k} N(k^\prime,t)$ $\sim
k^{1-\gamma}$, where $N(k^\prime,t)$ denotes the number of nodes
with degree $k^\prime$. The detailed analysis is given as follows.
For a degree $k$
\begin{equation}
k=q\left(\frac{[1+mp(q-1)]^{t-s}+q-2}{q-1}\right),
\end{equation}
there are  $L_v(s)=mp(q+1)(1+mpq)^{s-1}$ nodes with this exact
degree, all of which were born at step $s$. All nodes born at time
$s$ or earlier have this or a higher degree. So we have
\begin{eqnarray}
\sum_{k' \geq k}
N(k',t)=\sum_{a=0}^{s}L_v(a)=\frac{(q+1)[(mpq+1)^{s}+q-1]}{q}\nonumber.
\end{eqnarray}
As the total number of nodes at step $t$ is given in Eq.~(\ref{Nt1})
we have
\begin{eqnarray}
\left(\frac{[1+mp(q-1)]^{t-s}+q-2}{1-1/q}\right)^{1-\gamma}\nonumber
=\frac{\frac{(q+1)[(mpq+1)^{s}+q-1]}{q}}{\frac{(q+1)[(mpq+1)^{t}+q-1]}{q}}\nonumber.
\end{eqnarray}
Therefore, for large $t$ we obtain
\begin{equation}
\left[[1+mp(q-1)]^{t-s}\right]^{1-\gamma}=(1+mpq)^{s-t}
\end{equation}
and
\begin{equation}\label{gamma}
\gamma \approx 1+\frac{\ln (1+mpq)}{\ln[1+mp(q-1)]}.
\end{equation}
Thus, the degree exponent $\gamma$ is a continuous function of $m$,
$p$ and $q$, and belongs to the interval [2,3], coinciding with the
empirically found results. %For any fixed $q$, as $p$ decreases from
%1 to 0, $\gamma$ increases from $1+\frac{\ln (1+mq)}{\ln
%[1+m(q-1)]}$ to $2+\frac{1}{q-1}$ (see next subsubsection for the
%theoretic calculation of degree distribution for the particular case
%of $m=1$ and $p\rightarrow0$ but without reaching zero, by using a
%rate-equation approach). In the case $q=2$, $\gamma$ can be tunable
%between 2 and 3.
In some limiting cases, Eq. (\ref{gamma}) recovers the results
previously obtained in
Refs.~\cite{DoGoMe02,CoFeRa04,ZhRoZh07,DoMeSa01,Do03}.
Figure~\ref{Fig3} shows, on a logarithmic scale, the scaling
behavior of the cumulative degree distribution $P_{cum}(k)$ for
different values of $p$ in the case of $m=1$ and $q=2$. Simulation
results agree very well with the analytical ones.

%%%%%%%%%%%%%%%%%%%%%%%%%%%%%%%%%%%%%%%%%%%%%%%%%%%%%%%%%%
% Figure  2
%%%%%%%%%%%%%%%%%%%%%%%%%%%%%%%%%%%%%%%%%%%%%%%%%%%%%%%%%%
%%%%%%%%%%%%%%%%%%%%%%%%%%%%%%%%%%%%%%%%%%%%%%%%%%%%%%%%%%

\subsubsection{Rate equation for some limiting cases} When
$m=1$ and $p\rightarrow0$ (but $p \neq0$), our model turns out to be
the graph, which we call expanded random pseudofractal scale-free
network (ERPSFN) that evolves as follows (see~\cite{Do03} for
interpretation): starting with a ($q$+1)-clique ($t=0$), at each
time step, we choose an existing $q$-clique, then we add a new node
and join it to all the nodes of the selected $q$-clique. When $q$=
2, the particular model is exactly the random pseudofractal
scale-free network (RPSFN)~\cite{DoMeSa01}.

In fact, the expanded random pseudofractal scale-free network can be
easily mapped into the \emph{Yule process}~\cite{WiYu22,Yu25}, which
was inspired by observations of the statistics of biological taxa.
The Yule process can be prescribed mathematically as follows: we
measure the passage of time by the number of genera. At each time
step one new species founds a new genus, thereby increasing the
number of genera by 1, and $q$ other species are added to various
pre-existing genera which are selected in proportion to the number
of species they already have. Let the nodes and $q$-cliques in
ERPSFN correspond to genera and species, respectively, then the
mapping from ERPSFN to the Yule process is established. From this
perspective, our model may find some applications in biological
systems. Next, we show that the degree distribution of ERPSFN is
power-law with the same degree exponent as the Yule distribution.

Since the size of ERPSFN is incremented by one with each step, here
we use the step value $t$ to represent a node created at this step.
Furthermore, after a new node is added to the network, the number of
$q$-cliques increases by $q$. We can see easily that at step $t$,
the network consists of $N=t+q+1$ nodes and $N_{q}=qN-q^{2}+1$
$q$-cliques.

One can analyze the degree distribution mathematically as follows.
Given a node, when it is born, it has degree $q$, and the number of
$q$-cliques containing this node is also $q$. After that, when its
degree increases by one, the number of $q$-cliques with this node as
one of its components increases by $q-1$, so the number of
$q$-cliques for selection containing a node with degree $k$ is
$(q-1)k-q^{2}+2q$. We denote by $P_{k,N}$ the fraction of nodes with
degree $k$ when the network size is~$N$. Thus the number of such
nodes is~$NP_{k,N}$. Then the probability that the new node happens
to be connected to a particular node~$i$ having degree $k_i$ is
proportional to~$(q-1)k_i-q^{2}+2q$, and so when properly normalized
is just $[(q-1)k_i-q^{2}+2q]/(qN-q^{2}+1)$. Hence, between the
appearance of the $N$th and the $(N+1)$th node, the total expected
number of nodes with degree $k$ that gain a new link during this
interval is
\begin{equation}
{(q-1)k-q^{2}+2q\over qN-q^{2}+1} \times NP_{k,N} \simeq {q-1\over
q} kP_{k,N},
\end{equation}
which holds for large $N$. Observe that the number of nodes with
degree $k$ will decrease at each time step by exactly this number.
At the same time the number increases because of nodes that
previously had $k-1$ degrees and now have an extra one. Thus we can
write a rate equation~\cite{KaRe01} for the new number
$(N+1)P_{k,N+1}$ of nodes with degree $k$ as:
\begin{equation}
(N+1)P_{k,N+1} = NP_{k,N} + {q-1\over q}
                 \left[ (k-1)P_{k-1,N} - k P_{k,N} \right].
\label{yule1}
\end{equation}
The only exception to Eq.~(\ref{yule1}) is for nodes having degree
$q$, which instead obey the equation
\begin{equation}
(N+1)P_{q,N+1} = NP_{q,N} + 1 - {q-1\over q}q P_{q,N}, \label{yule2}
\end{equation}
since by construction exactly one new such node appears at each time
step. When $N$ approaches infinity, we assume that the degree
distribution tends to some fixed value $P_k = \lim_{N\to\infty}
P_{N,k}$. Then from Eq.~(\ref{yule2}), we have
\begin{equation}
P_q = 1/q. \label{p1}
\end{equation}
And Eq.~(\ref{yule1}) becomes
\begin{equation}
P_k = {q-1\over q} \left[ (k-1) P_{k-1} - k P_k \right],
\end{equation}
from which we can easily obtain the recursive equation
\begin{equation}
P_k = {k-1\over k+1+\frac{1}{q-1}}\,P_{k-1},
\end{equation}
which can be iterated to get
\begin{eqnarray}
P_k &=& {(k-1)(k-2)\ldots q\over(k+1+\frac{1}{q-1})(k+\frac{1}{q-1})\ldots(q+2+\frac{1}{q-1})}\,P_q\nonumber\\
    &=& {(k-1)(k-2) \ldots (q+1)\over (k+1+\frac{1}{q-1})(k+\frac{1}{q-1})\ldots(q+2+\frac{1}{q-1})},
\end{eqnarray}
where Eq.~(\ref{p1}) has been used.  This can be simplified further
by making use of a handy property of the $\Gamma$-function,
$\Gamma(a)=(a-1)\Gamma(a-1)$ with $\Gamma(a)$ defined by:
\begin{equation}
\Gamma(a) = \int_0^\infty x^{a-1} e ^{-x} d x. \label{defsgamma}
\end{equation}  By this property and $\Gamma(1)=1$, we get{\footnotesize
\begin{eqnarray}\label{simon}
P_k &=& \frac{(q+1+\frac{1}{q-1})(q+\frac{1}{q-1})\ldots (2+\frac{1}{q-1})}{q(q-1)\ldots 1} {\Gamma(k)\Gamma(2+\frac{1}{q-1})\over\Gamma(k+2+\frac{1}{q-1})} \nonumber\\
    &=& \frac{(q+1+\frac{1}{q-1})(q+\frac{1}{q-1})\ldots (2+\frac{1}{q-1})}{q(q-1)\ldots 1}\mathbf{B}(k,2+\frac{1}{q-1}),\nonumber\\
\label{gammas}
\end{eqnarray}}
where $\mathbf{B}(a,b)$ is the Legendre beta-function, which is
defined as
\begin{equation}
\textbf{B}(a,b) = {\Gamma(a)\Gamma(b)\over\Gamma(a+b)},
\label{defsbeta}
\end{equation}
 Note that the beta-function has the interesting
property that for large values of either of its arguments it itself
follows a power law. For instance, for large $a$ and fixed~$b$,
$\textbf{B}(a,b)\sim a^{-b}$. Then we can immediately see that for
large $k$, $P_k$ also has a power-law tail with a degree exponent
\begin{equation}
\gamma = 2 + {1\over q-1}. \label{yuleexponent}
\end{equation}
For $q=2$,  $\gamma=3$, which has been obtained previously
in~\cite{Do03}. Note that Eq.~(\ref{simon}) is similar to the Yule
distribution~\cite{Ne05} called by Simon~\cite{Simon55}. %In order to
%confirm the validity of the obtained analytical degree exponent of
%ERPSFN, we perform numerical simulations reported in
%Fig.~\ref{Yule}, which shows that numerical and analytical results
%are in agreement with each other.

%%%%%%%%%%%%%%%%%%%%%%%%%%%%%%%%%%%%%%%%%%%%%%%%%%%%%%%%%%
% Figure  3
%%%%%%%%%%%%%%%%%%%%%%%%%%%%%%%%%%%%%%%%%%%%%%%%%%%%%%%%%%
%\begin{figure}
%\begin{center}
%\begin{tabular}{cc}
        %\includegraphics[width=4cm]{degreeq2.eps}
        %\includegraphics[width=4cm]{degreeq3.eps}
        %\includegraphics[width=7cm]{pic/net4-2.2.eps}&\includegraphics[width=7cm]{pic/net4-3.3.eps}
   % \end{tabular}
%\includegraphics[width=9cm]{degree.eps}
%\caption{The cumulative degree distribution $P_{cum}(k)$ of the
%ERPSFN with size 150000 for the case of $q=2$ and $q=3$. The dashed
%lines are the theoretical results of $\gamma(q)$ shown in Eq.
%(\ref{yuleexponent}). All simulation data are from the average of 50
%independent runs.} \label{Yule}
%\end{center}
%\end{figure}
%%%%%%%%%%%%%%%%%%%%%%%%%%%%%%%%%%%%%%%%%%%%%%%%%%%%%%%%%%

\subsection{Clustering coefficient}

In a network if a given node is connected to $k$ nodes, defined as
the neighbors of the given node, then the ratio between the number
of links among its neighbors and the maximum possible value of such
links $k(k-1)/2$ is the clustering coefficient of the given
node~\cite{WaSt98}. The clustering coefficient of the whole network
is the average of this coefficient over all nodes in the network.

For our network, the analytical expression of clustering coefficient
$C(k)$ for a single node with degree $k$ can be derived exactly.
When a node is created it is connected to all the nodes of a
$q$-clique, in which nodes are completely interconnected. So its
degree and clustering coefficient are $q$ and 1, respectively. In
the following steps, if its degree increases one by a newly created
node connecting to it, then there must be $q-1$ existing neighbors
of it attaching to the new node at the same time. Thus for a node of
degree $k$, we have
\begin{equation}\label{Ck}
C(k)= {{{q(q-1)\over 2}+ (q-1)(k-q)} \over {k(k-1)\over 2}}=
\frac{2(q-1)(k-\frac{q}{2})}{k(k-1)},
\end{equation}
which depends on both $k$ and $q$. For $k \gg q$, the $C(k)$ is
inversely proportional to degree $k$. The scaling $C(k)\sim k^{-1}$
has been found for some network
models~\cite{AnHeAnSi05,DoMa05,ZhCoFeRo05,ZhYaWa05,ZhRoCo05,ZhRoZh06,DoGoMe02,CoFeRa04,ZhRoZh07,DoMeSa01,Do03,RaBa03},
and has also been observed in several real-life
networks~\cite{RaBa03}.

Using Eq. (\ref{Ck}), we can obtain the clustering $\overline{C}_t$
of the networks at step $t$:
\begin{equation}\label{ACCk}
\overline{C}_t=
    \frac{1}{N_{t}}\sum_{r=0}^{t}
    \frac{2(q-1)(D_r-\frac{q}{2})L_v(r)}{D_r(D_r-1)},
\end{equation}
where the sum runs over all the nodes
and $D_r$ %=(d+1)\left(\frac{[1+(d-1)q]^{t-r}+d-2}{d-1}\right)$
is the degree of the nodes created at step $r$, which is given by
Eq. (\ref{Ki1}).

%%%%%%%%%%%%%%%%%%%%%%%%%%%%%%%%%%%%%%%%%%%%%%%%%%%%%%%%%%
% Figure  3
%%%%%%%%%%%%%%%%%%%%%%%%%%%%%%%%%%%%%%%%%%%%%%%%%%%%%%%%%%
\begin{figure}
\begin{center}
\includegraphics[width=8cm]{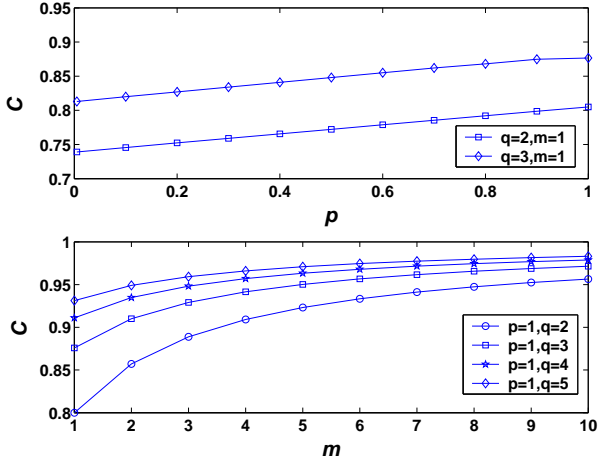}
\caption{The dependence relation of network clustering coefficient
$C$ on $m$, $p$, and $q$. Results are averaged over ten network
realizations for each datum.} \label{ACC1}
\end{center}
\end{figure}
%%%%%%%%%%%%%%%%%%%%%%%%%%%%%%%%%%%%%%%%%%%%%%%%%%%%%%%%%%

In the infinite network order limit ($N_{t}\rightarrow \infty$), Eq.
(\ref{ACCk}) converges to a nonzero value $C$. Obviously, network
clustering coefficient $\overline{C}_t$ is a function of parameters
$m$, $p$ and $q$. If we fixed any two of them, $\overline{C}_t$
increases with the rest. Exactly analytical computations show: in
the case $m=1$ and $q=2$, when $p$ increases from 0 to 1, $C$ grows
from 0.739 \cite{BoPa05} to 0.8 \cite{DoGoMe02}; In the case $p=1$
and $q=2$, when $m$ increases from 1 to infinite, $C$ grows from 0.8
\cite{DoGoMe02} to 1; Likewise, in the case $m=1$ and $p=1$, $C$
increases from 0.8 to 1 when $q$ increases from 2 to infinite, with
special values $C=0.8571$ and $C=0.8889$ for $q=3$ and $q=4$,
respectively. Therefore, the average clustering coefficient is very
large, which shows the evolving networks are highly clustered.
Figure~\ref{ACC1} exhibits the dependence of the clustering
coefficient $C$ on $m$, $p$ and $q$, which agree well with our above
conclusions.

From Figs.~\ref{Fig3} and~\ref{ACC1} and Eqs. (\ref{gamma}) and
(\ref{ACCk}), one can see that both degree exponent $\gamma$ and
clustering coefficient $C$ depend on the parameter $m$, $p$, and
$q$. The mechanism resulting in this relation should be paid further
effort. The fact that there is a biased choice of the cliques at
each evolving step may be a possible explanation, see
Ref.~\cite{CoRobA05}.

\subsection{Average path length}
The most important property for a small-world network is a
logarithmic average path length (APL) (with the number of nodes). It
has obvious implications for the dynamics of processes taking place
on networks. Therefore, its study has attracted much attention. Here
APL means the minimum number of edges connecting a pair of nodes,
averaged over all pairs of nodes. In this subsection, first we give
an upper bound of APL for the general case; then, we compute exactly
the APL for a particular deterministic network. Both of the obtained
values grow logarithmically with the network size.

\subsubsection{An upper bound of APL for general case}
We denote the network nodes by the time step of their generations,
$v=1,2,3,\cdots,N-1,N$. Using $L(N)$ to represent the APL of the our
model with system size $N$, then we have following relation:
$L(N)=\frac{2\sigma(N)}{N(N-1)}$, where $\sigma(N)=\sum_{1 \leq i<j
\leq N}d_{i,j}$ is the total distance, in which $d_{i,j}$ is the
shortest distance between node $i$ and node $j$. By using the
approach similar to that
in~\cite{ZhRoCo05a,ZhYaWa05,ZhRoCo05,ZhRoZh06}, we can evaluate the
APL of the present model.

Obviously, when a new node enters the networks, the smallest
distances between existing node pairs will not change. Hence we have
\begin{equation}\label{E6}
\sigma(N+1) = \sigma(N)+ \sum_{i=1}^{N}d_{i,N+1}.
\end{equation}
Equation~(\ref{E6}) can be approximately represented as:
\begin{equation}\label{E7}
\sigma(N+1) = \sigma(N)+N+(N-q)L(N-q+1),
\end{equation}
where
\begin{equation}\label{E8}
(N-q)L(N-q+1) = {2\sigma(N-q+1) \over N-q+1} < {2\sigma(N) \over N}.
\end{equation}
Equations~(\ref{E7}) and~(\ref{E8}) provide an upper bound for the
variation of $\sigma(N)$ as
\begin{equation}\label{E9}
{d\sigma(N) \over dN} =  N + {2\sigma(N) \over N},
\end{equation}
which yields
\begin{equation}
\sigma(N) = N^2(\ln N + \omega),
\end{equation}
where $\omega$ is a constant. As $\sigma(N) \sim N^2\ln N $, we have
$L(N) \sim \ln N$.

%%%%%%%%%%%%%%%%%%%%%%%%%%%%%%%%%%%%%%%%%%%%%%%%%%%%%%%%%%
% Figure  4
%%%%%%%%%%%%%%%%%%%%%%%%%%%%%%%%%%%%%%%%%%%%%%%%%%%%%%%%%%
\begin{figure}
\begin{center}
\includegraphics[width=8cm]{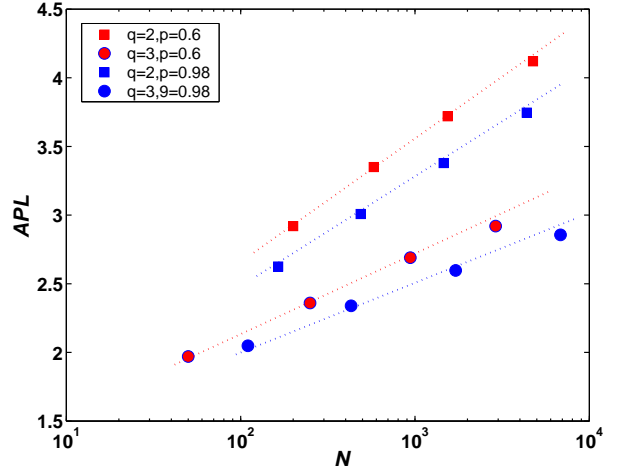} \caption{Semilogarithmic graph of the APL vs
the network size $N$ in the special case of $m=1$. Each datum point
is obtained as an average of 50 independent network realizations.
The lines are linear functions of $\ln N$.} \label{Fig4}
\end{center}
\end{figure}
%%%%%%%%%%%%%%%%%%%%%%%%%%%%%%%%%%%%%%%%%%%%%%%%%%%%%%%%%%

Note that Eq.~(\ref{E9}) was deduced from an inequality, which
implies that the increasing tendency of $L(N)$  is at most as $\ln
N$ with $N$. Thus, our model exhibits the presence of small-world
property. In Fig.~\ref{Fig4}, we show the dependence of the APL on
system size $N$ for different $p$ and $q$ in the case of $m=1$. From
Fig.~\ref{Fig4}, one can see that for fixed $q$, APL decreases with
increasing $p$; and for fixed $p$, APL is a decreasing function of
$q$. When network size $N$ is small, APL is a linear function of
$\ln N$; while $N$ becomes large, APL increases slightly slower than
$\ln N$. So the simulation results are in agreement with the
analytical prediction. It should be noted that in our model, if we
fix  $p$ and $q$, considering other values of $m$ greater than 1,
then the APL will increase more slowly than in the case $m = 1$ as
in those cases the larger $m$ is, the denser the network becomes.

\subsubsection{Exact result of APL for a special case}
 For $p=1$, the networks
are deterministic, which allows one to calculate the APL
analytically. Here we only consider a particular case of $m=1$,
$p=1$, and $q=2$, which we denote after $t$ generations by $Q_{t}$.
Then the average path length of $Q_t$ is defined to be:
\begin{equation}\label{eq:app4}
  \bar{d}_t  = \frac{D_t}{N_t(N_t-1)/2}\,,
\end{equation}
where
\begin{equation}\label{eq:app5}
  D_t = \sum_{i,j \in Q_t} d_{i,j}\,.
\end{equation}

\begin{figure}[t]
  \centering\includegraphics[width=7cm]{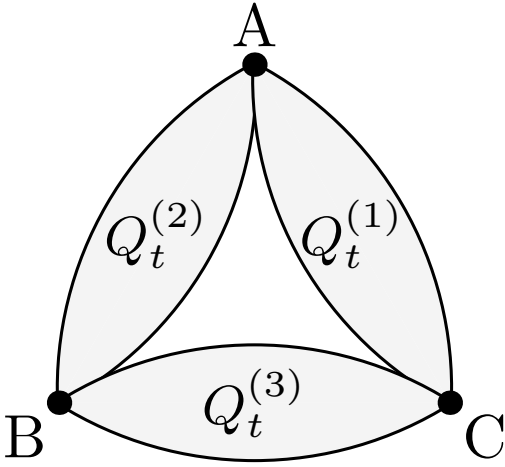}
  \caption{The network after $t+1$
    generations, $Q_{t+1}$, is obtained by joining three copies of generation $t$
    (i.e. $Q_t^{(1)}, Q_t^{(2)}, Q_t^{(3)}$) at the three nodes of highest degree, denoted by $A$, $B$, $C$.}\label{apfig2}
\end{figure}

The deterministic recursive construction of this particular network
has a self-similar structure that allows us to exactly calculate
$\bar{d}_t$ by following a similar approach introduced
in~\cite{HiBe06}. As shown in Fig.~\ref{apfig2}, the network
$Q_{t+1}$ may be obtained by joining at the hubs (the most connected
nodes) three copies of $Q_t$, which we label $Q_t^{(\alpha)}$,
$\alpha=1,2,3$~\cite{Bobe05}. Then one can write the sum over all
shortest paths $D_{t+1}$ as
\begin{equation}\label{eq:app6}
  D_{t+1} = 3D_t + \Delta_t\,,
\end{equation}
where $\Delta_t$ is the sum over all shortest paths whose endpoints
are not in the same $Q_t$ branch. The solution of
Eq.~(\ref{eq:app6}) is
\begin{equation}\label{eq:app8}
  D_t = 3^{t-1} D_1 + \sum_{\tau=1}^{t-1} 3^{t-\tau-1} \Delta_\tau\,.
\end{equation}
The paths that contribute to $\Delta_\tau$ must all go through at
least one of the three hubs ($A$, $B$, $C$) where the three
different $Q_t$ branches are joined. Below we give the analytical
expression for $\Delta_t$ named the crossing paths, which is given
by
\begin{equation}\label{APL31}
\Delta_t=\Delta_t^{1,2} + \Delta_t^{2,3} + \Delta_t^{1,3}\,,
\end{equation}
where $\Delta_t^{\alpha,\beta}$ denotes the sum of all shortest
paths with endpoints in $Q_t^{(\alpha)}$ and $Q_t^{(\beta)}$. It
should be noted that $\Delta_t^{\alpha,\beta}$ excludes the paths
where either endpoint is the hub they have in common, and includes
only one of the paths from the unshared hub in one $Q_t$ branch
(e.g. $Q_t^{(\alpha)}$) to all non-hub nodes in the other $Q_t$
branch (e.g. $Q_t^{(\beta)}$).

By symmetry, $\Delta_t^{1,2} = \Delta_t^{2,3} = \Delta_t^{1,3}$, we
have
\begin{equation}\label{APL32}
\Delta_t = 3 \Delta_t^{1,2}\,,
\end{equation}
where $\Delta_t^{1,2}$ is given by the sum
\begin{align}\label{app12}
  \Delta_t^{1,2} &= \sum_{\substack{i \in Q_t^{(1)},\,\,j\in
      Q_t^{(2)}\\ i \ne A,C,\,\,j\ne A}} d_{i,j}
\end{align}
In order to find $\Delta_t^{1,2}$, we define
\begin{align}
d_t^\text{tot} &\equiv \sum_{Z \in Q_t^{(1)}}d_{Z,A}\, ,\nonumber\\
d_t^\text{near} &\equiv \sum_{\substack{Z \in Q_t^{(1)}\\ d_{Z,A} <
    d_{Z,C}}} d_{Z,A}\,,\qquad N_t^\text{near} \equiv \sum_{\substack{Z
    \in Q_t^{(1)}\\ d_{Z,A} < d_{Z,C}}} 1\,,\nonumber\\
d_t^\text{mid} &\equiv \sum_{\substack{Z \in Q_t^{(1)}\\ d_{Z,A} =
    d_{Z,C}}} d_{Z,A}\,,\qquad N_t^\text{mid} \equiv \sum_{\substack{Z
    \in Q_t^{(1)}\\ d_{Z,A} = d_{Z,C}}} 1\,,\nonumber\\
d_t^\text{far} &\equiv \sum_{\substack{Z \in Q_t^{(1)}\\ d_{Z,A} >
    d_{Z,C}}} d_{Z,A}\,,\qquad N_t^\text{far} \equiv \sum_{\substack{Z
    \in Q_t^{(1)}\\ d_{Z,A} > d_{Z,C}}} 1\,,
\label{eq:app32}
\end{align}
where $Z\neq A$ and $Z\neq C$. Then we can easily have
$d_t^\text{tot} = d_t^\text{near} + d_t^\text{mid} + d_t^\text{far}$
and $N_t = N_t^\text{near} + N_t^\text{mid} + N_t^\text{far} + 2$.
By symmetry $N_t^\text{near} = N_t^\text{far}$. Thus, by
construction, we obtain
\begin{equation}
\left\{\begin{array}{lc} {\displaystyle {N_t= 2\,N_t^\text{near} +
N_t^\text{mid} +2}},\\
{\displaystyle N_{t+1}^\text{mid}=N_t^\text{mid}+
2\,N_t^\text{near}+1.}
\end{array} \right.
\end{equation}
Using these two relations and considering the initial values, we
obtain partial quantities in Eq.~\eqref{eq:app32} as
\begin{equation}\label{APL36}
 \left\{\begin{array}{lc}
  N_t^\text{far} = N^\text{near}_t = \frac{1}{6}\left(-3+3^{t+1}\right),\\
  N^\text{mid}_t = \frac{1}{6} \left(3+3^{t+1}\right).
  \end{array} \right.
\end{equation}
Now we return to the quantity $\Delta_t^{1,2}$ which can be further
decomposed into the sum of four terms as
\begin{align}
\Delta_t^{1,2} &= \sum_{\substack{i \in Q_t^{(1)},\,\,j\in
      Q_t^{(2)}\\ i \ne A,C,\,\,j\ne A}} d_{i,j} \nonumber\\
&= \sum_{\substack{i \in Q_t^{(1)},\,\,j\in
      Q_t^{(2)},\,\,i \ne A,C,\,\,j\ne A,B \\ d_{i,A} > d_{i,C},\,\, d_{j,A} > d_{j,B}}}
  (d_{i,C}+d_{j,B}+1)\nonumber\\
&\quad   + \sum_{\substack{i \in Q_t^{(1)},\,\,j\in
      Q_t^{(2)} \\i \ne A,C,\,\,j\ne A,B,\,\, d_{i,A}\le d_{i,C}}}
  (d_{i,A}+d_{A,j})\nonumber \\ &\quad + \sum_{\substack{i \in Q_t^{(1)},\,\,j\in
      Q_t^{(2)},\,\,i \ne A,C,\,\,j\ne A,B \\d_{i,A} > d_{i,C},\,\,d_{j,A}\le d_{j,B}}}
  (d_{i,A}+d_{A,j})\nonumber\\ &\quad  + \sum_{\substack{i \in Q_t^{(1)},\,\,i \ne A,C,\,\,j=B}}
  (d_{i,A}+1)\nonumber\\
&= 2N_t^\text{near} d_t^\text{near}+(N_t^\text{near})^2\nonumber\\
&\quad +(N_t^\text{near}+N_t^\text{mid})(d_t^\text{near}+d_t^\text{mid}+d_t^\text{far})\nonumber\\
&\quad +(N_t-2)(d_t^\text{near}+d_t^\text{mid})\nonumber\\
&\quad +(N_t^\text{near}+N_t^\text{mid})d_t^\text{far}\nonumber\\
&\quad +N_t^\text{near}(d_t^\text{near}+d_t^\text{mid})\nonumber\\
&\quad +d_t^\text{near}+d_t^\text{mid}+d_t^\text{far}+N_t-2\,.
\label{APL37}
\end{align}
Having $\Delta_n^{1,2}$ in terms of the quantities in
Eq.~\eqref{eq:app32}, the next step is to explicitly determine these
quantities unresolved.

Since $A$ and $C$ are linked by one edge, for any node $i$ in the
network, $d_{i,A}$ and $d_{i,C}$ can differ by at most 1. In
addition, considering the self-similar structure of the graph, we
can easily know that at time $t+1$, the quantities
$d_{t+1}^\text{mid}$, $d_{t+1}^\text{near}$ and $d_{t+1}^\text{far}$
evolve as
\begin{equation}%\label{eq5}
\left\{\begin{array}{lc} {\displaystyle{d^\text{mid}_{t+1} =
d^\text{mid}_t+2\,d^\text{far}_t+1\,,} }\\
{\displaystyle{d^\text{near}_{t+1} =
d^\text{mid}_t+2\,d^\text{near}_t\, ,} }\\
{\displaystyle{d^\text{far}_{t+1} = d^\text{mid}_t+2\,d^\text{far}_t+N_t^\text{mid}\,.} }\\
\end{array} \right.
\end{equation}
From these recursive equations we can obtain
\begin{equation}\label{APL39}
\left\{\begin{array}{lc} {\displaystyle{d^\text{mid}_{t} =
3^{t-2}\,(t+5)\,,} }\\
{\displaystyle{d^\text{near}_{t} = 3^{t-2}\, (t+2)\, ,} }\\
{\displaystyle{d^\text{far}_{t} = \frac{1}{54}
\left(2\,(t+1)\cdot3^{t+1}
   \,+11\cdot3^{t+1}-27\right)\,.} }\\
\end{array} \right.
\end{equation}
Substituting the obtained expressions in Eqs.~\eqref{APL36} and
\eqref{APL39} into Eqs.~\eqref{APL37} and~\eqref{APL32}, the
crossing paths $\Delta_t$ is obtained to be
\begin{equation}\label{APL10}
  \Delta_t = \frac{1}{12}\left[(4\,t+13)\,9^{t}-9\right].
\end{equation}
Inserting Eq.~\eqref{APL10} into Eq.~\eqref{eq:app8} and using $D_1
= 21$, we have
\begin{equation}\label{APL11}
  D_t = \frac{1}{8} \left( 4\,t\cdot
  9^t+10\cdot3^t+11\cdot9^t+3\right).
\end{equation}
Substituting Eqs. (\ref{Nt1}) and (\ref{APL11}) into
(\ref{eq:app4}), the exact expression for the average path length is
obtained to be
\begin{equation}\label{eq:app10}
  \bar{d}_t = \frac{ 4\,t\cdot
  9^t+10\cdot3^t+11\cdot9^t+3}{3+4\cdot3^{t+1}+9^{t+1}}.
\end{equation}
In the infinite network size limit ($t \rightarrow \infty$),
\begin{equation}\label{eqapp56}
\bar{d}_{t} \simeq \frac{4}{9} t+\frac{11}{9} \sim \ln N_{t},
\end{equation}
which means that the average path length shows a logarithmic scaling
with the size of the network.

 It should be mentioned that the final
expressions contained in Eqs.~\eqref{eq:app10} and~\eqref{eqapp56}
were quoted earlier in Ref.~\cite{DoGoMe02} [Eqs. (6) and (7) of
that work]. However Ref.~\cite{DoGoMe02} did not provide any of the
details of the derivation, so the explicit calculation presented
here is pedagogically useful. Moreover, the analytical method may
guide and shed light on related studies for other deterministic
network models.

\section{Conclusions and discussions}

In summary, we have proposed and studied a class of evolving
networks consisting of cliques. We have obtained the analytical and
numerical results for degree distribution, clustering coefficient,
as well as the average path length, which are determined by the
model parameters and in accordance with large amount of real
observations. The networks are scale-free, with degree exponent
adjusted continuously between 2 and 3. The clustering coefficient of
single nodes has a power-law spectra, the network clustering
coefficient is very large and independent of network size. The
intervertex separation is small, which increases at most
logarithmically as the network size.

In real-life world, many networks consist of cliques. For example,
in movie actor collaboration network~\cite{WaSt98} and science
collaborating graph~\cite{Ne01a}, actors acting in the same film or
authors signing in the same paper form a clique, respectively. In
corporate director network~\cite{BaCa04}, directors as members in
the same board constitute a clique. Analogously, in public transport
networks~\cite{SiHo05}, bus (tramway, or underground) stops shape a
clique if they are consecutive stops on a route, and in the network
of concepts in written texts~\cite{CaLoAnNeMi06}, words in each
sentence in the text are added to the network as a clique. All these
pose a very interesting and important question of how to build
evolution models based on this particularity of network
component---cliques. Interestingly, our networks, although different
from real world, are formed by cliques, this particularity of the
composing units may provide a comprehensive aspect to understand
some real-life systems. In future, it would be more interesting to
establish a model describing real systems consisting of cliques such
as actor collaboration network where cliques arise from mutual
cooperation~\cite{GuLa04,GuLa06}.

Future work should also include studying in detail dynamical
processes taking place on our networks, which may provide some
original and interesting results to the field. For example, it
should be possible to adapt the renormalization-group techniques
used in Refs.~\cite{HiBe06,ErTuYuBe05,Hi07} for studying the Ising
model to examine cooperative behavior on the general case of our
model. There could be a variety of interesting phase transition
behaviors as the network structure is modified through the
parameters $m$, $p$,
 and $q$.

\section*{Acknowledgment}
The authors would like to thank Zhen Shen for his assistance in
preparing the manuscript.  This research was supported by the
National Basic Research Program of China under grant No.
2007CB310806, the National Natural Science Foundation of China under
Grant Nos. 60496327, 60573183, and 90612007, the Postdoctoral
Science Foundation of China under Grant No. 20060400162, the Program
for New Century Excellent Talents in University of China
(NCET-06-0376), and the Huawei Foundation of Science and Technology
(YJCB2007031IN).

%%%%%%%%%%%%%%%%%%%%%%%%%%%%%%%%%%%%%%%%%%%%%%%%%%%%%%%%%%%%%%%%%
%%%%%%%%%%%%%%%%%%%%%%%%%%%%%%%%%%%%%%%%%%%%%%%%%%%%%%%%%%%%%%%%%

%\section{References}

\end{document}